\documentclass[preprint2]{emulateapj}
\usepackage{color}
\usepackage{hyperref}
\usepackage{multirow}
\usepackage{amsmath}
\usepackage{tabularx}
\slugcomment{Published in The Astrophysical Journal 742 (2011) 2; submitted 2011 March 1; accepted 2011 August 12}
\shorttitle{PAHs and interstellar UV-VIS absorption bands}
\shortauthors{Steglich et al.}

\begin{document}
\title{Can Neutral and Ionized Polycyclic Aromatic Hydrocarbons be Carriers of the\\ Ultraviolet Extinction Bump and the Diffuse Interstellar Bands?}
\author{M. Steglich$^1$, J. Bouwman$^{2,3}$, F. Huisken$^1$, and Th. Henning$^4$}
\affil{$^1$ Laboratory Astrophysics Group of the Max Planck Institute for Astronomy at the Friedrich Schiller University Jena,\\
Institute of Solid State Physics, Helmholtzweg 3, D-07743 Jena, Germany; M.Steglich@web.de\\
$^2$ Raymond and Beverly Sackler Laboratory for Astrophysics, Leiden Observatory, Leiden University,\\
P.O. Box 9513, NL 2300 RA Leiden, The Netherlands\\
$^3$ Present address: Departments of Chemistry and Physics, University of California, Berkeley, California 94720, USA\\
$^4$ Max Planck Institute for Astronomy, K\"onigstuhl 17, D-69117 Heidelberg, Germany}
%\email{}
%\altaffiltext{4}{}

\begin{abstract}
Up to now, no laboratory-based study has investigated polycyclic aromatic hydrocarbon (PAH) species as potential carriers of both the diffuse interstellar bands (DIBs) and the 2175 \AA\ UV bump. We examined the proposed correlation between these two features by applying experimental and theoretical techniques on two specific medium-sized/large PAHs (dibenzorubicene C$_{30}$H$_{14}$ and hexabenzocoronene C$_{42}$H$_{18}$) in their neutral and cationic states. It was already shown that mixtures of sufficiently large, neutral PAHs can partly or even completely account for the UV bump. We investigated how the absorption bands are altered upon ionization of these molecules by interstellar UV photons. The experimental studies presented here were realized by performing matrix isolation spectroscopy with subsequent far-UV irradiation. The main effects were found to be a broadening of the absorption bands in the UV combined with slight redshifts. The position of the complete $\pi$ -- $\pi^*$ absorption structure around 217.5 nm, however, remains more or less unchanged which could explain the observed position invariance of the interstellar bump for different lines of sight. This favors the assignment of this feature to the interstellar PAH population. As far as the DIBs are concerned, neither our investigations nor the laboratory studies carried out by other research groups support a possible connection with this class of molecules. Instead, there are reasonable arguments that neutral and singly ionized cationic PAHs cannot be made responsible for the DIBs.
\end{abstract}


\section{Introduction}
Mid-infrared (MIR) emission bands related to aromatic C--H and C--C vibrational modes were found in several astrophysical environments, ranging from planetary nebula, reflection nebulosities, circumstellar disks to even active galactic nuclei \citep[see, e.g., the recent review of][]{tielens08}. Inferred from their presence, polycyclic aromatic hydrocarbons (PAHs) are believed to be an important component of interstellar dust, and, after H$_2$ and CO, they are probably the most abundant molecules in the interstellar medium \citep[ISM; see, e.g.,][]{leger84}. Although these MIR bands do not permit an identification of specific PAH molecules, general information, such as size and charge state distribution or the aliphatic-to-aromatic ratio, can be inferred \citep{draine07}. Based on energetic arguments, the MIR emission bands were attributed to aromatic molecules containing about (or at least) 50 carbon atoms \citep{tielens08}. Besides the elaborately investigated IR properties of PAHs, it is equally important to perform experimental studies in other wavelength regimes to test whether the PAH model is compatible with observations. Usually, neutral or ionized PAHs possess no or only weak permanent dipole moments. Therefore, strong spectroscopic fingerprints due to pure rotational transitions in the radio range are not expected. On the other hand, electronic $\pi$ -- $\pi^*$ transitions give rise to distinct absorption bands in the UV-VIS. The ISM displays two clear spectroscopic features in this wavelength range: a growing number of yet unidentified diffuse interstellar bands (DIBs) above 400 nm and a broad bump at 217.5 nm with almost fixed position, but varying strength and width for different lines of sight. These features are superimposed on a background with increasing extinction at shorter wavelengths \citep[see, e.g.,][]{draine03}. Below 400 nm, the extinction curve is smooth and no narrow bands that can be related to polyatomic gas phase molecules were found \citep{clayton03, gredel11, salama11}. Unfortunately, electronic spectra of astrophysically relevant \textit{large} PAHs and mixtures of them measured in the laboratory under appropriate conditions are almost unavailable, mainly because sufficient quantities of suitable samples are difficult to obtain. \citet{ruiterkamp02} and \citet{halasinski03} studied the UV-VIS absorption properties of a few selected matrix-isolated PAHs consisting of max. 48 C atoms. These studies were guided by the search for possible DIB carriers and were biased toward more exotic or elongated structures because of their strong absorption bands in the VIS. By measuring the absorbance of mixtures of hot gaseous PAHs with mean sizes around 31 C atoms, \citet{joblin92} found that the strongest spectral feature is a broad bump at 210 nm, very close and similar in shape to the interstellar extinction bump at 217.5 nm. Recently, we obtained analogous results for mixtures of comparably dimensioned PAHs which were isolated in solid Ne at low temperature \citep{steglich10}. A strong UV bump was found around 200 nm accompanied by a rather featureless extinction curve if molecules with less than 22 C atoms were absent from the mixture. Considering semi-empirical calculations, we argued that the bump position may shift to the red, eventually reaching 217.5 nm, for species large enough to be associated with the MIR emission bands ($\gtrsim$ 50 C atoms). This idea is furthermore supported by calculations applying time-dependent density functional theory (TD-DFT). \citet{malloci04} showed that a PAH mixture, comprising a limited number of different structures, could also give rise to a bump around 217.5 nm.

While more than 400 DIBs are known today, none of their carriers which are thought to be large carbon-bearing molecules in the gas phase, such as PAHs and their ions, have been identified yet \citep[for a review on DIBs see][]{sarre06}. An unambiguous identification requires the comparison with experimental spectra of the species under consideration. Usually, this demands for the application of elaborate experimental techniques, like cavity ring down spectroscopy carried out at low temperature in the gas phase. So far, the spectra of only a few small PAHs consisting of less than 25 C atoms have been recorded with this method \citep[for a review see][]{salama08}. When in their neutral charge state, such small molecules display their strongest absorption bands in the UV. As already mentioned, no sharp absorption bands were found in the UV part of the interstellar extinction curve implying a very small abundance of small neutral PAHs \citep{clayton03, gredel11, salama11}. The cations, on the other hand, show additional strong bands in the VIS wavelength range. However, these bands are usually too broad compared to the widths of most DIBs. Recent attempts to search for bands of the naphthalene (C$_{10}$H$_{8}$) and anthracene (C$_{14}$H$_{10}$) cations in interstellar spectra were proven to be unsuccessful \citep{galazutdinov11, searles11}.

This article presents new experimental results on the UV-VIS absorption properties of medium-sized to large PAHs. Supported by theoretical calculations, we attempt to solidify the connection between large PAHs and the UV bump. Bearing in mind that, for different lines of sight, the molecular population is affected by different physical conditions, like the strength of the UV field, it is important to address the question whether ionization has an effect on the bump position of PAH mixtures. We approach this issue experimentally through photoionization of two matrix-isolated PAHs, dibenzorubicene C$_{30}$H$_{14}$ (DBR) and hexabenzocoronene C$_{42}$H$_{18}$ (HBC), which display strong absorption bands around 217.5 nm in their neutral state. For the first time, complete electronic spectra of PAH ions spanning the full UV-VIS range down to 200 nm will be presented. The spectrum of the HBC cation will be of particular interest as, in addition to its superior stability \citep{troy06}, this molecule has a size comparable to the PAH size that was inferred from the aromatic IR emission bands.

\section{Predictions inferred from time-dependent density functional theory} \label{tddft}
\subsection{UV Bump at 217.5 nm} \label{bump_sec}
In this section, we shortly address the problem of the UV signature of PAHs (the bump below 230 nm) and its size dependence using computational methods. Previously, we already calculated the absorbance of artificial PAH mixtures with different mean sizes by applying a semi-empirical method \citep{steglich10}. The low computational costs of this method, however, may be accompanied by rather unprecise band positions and strengths. Furthermore, size-dependent effects could partially be related to the model itself. This can be avoided by using more elaborate ab initio and TD-DFT methods which, on the other hand, can only be applied to a limited number of molecules. \citet{malloci04} calculated the absorption spectra of 40 PAHs in different charge states using a special treatment of TD-DFT implemented in the Octopus software package \citep{castro06,marques04}. These spectra were later collected in an online database and used by \citet{cecchi08} to simulate artificial extinction curves for a dust model that includes PAHs. In this model, the UV bump and the non-linear far-UV (FUV) rise of the interstellar extinction curve were solely attributed to electronic transitions in PAHs. %However, the molecules that were included spanned a large size range from 10 to 66 C atoms. Laboratory studies on a real mixture of PAHs containing a significant amount of small PAHs ($\lesssim$ 22 C atoms) revealed additional structures on the UV part of the extinction curve \citep[see][]{steglich10} that are not observed in the ISM. Besides, the Octopus treatment of TD-DFT generates PAH spectra which, when compared with laboratory measurements, are not very accurate for wavelengths longer than 200 nm. Especially, the electronic transitions above 230 nm are usually redshifted and too weak compared to the $\pi$ -- $\pi^*$ and $\sigma$ -- $\sigma^*$ transitions responsible for the UV bump and the FUV rise.

Here, we used the TD-DFT implementation of the Gaussian09 software \citep{frisch09} which relies on basis sets to create the molecular orbitals. The computational effort of this method scales steeply with the number of electronic transitions to be calculated and, therefore, it is much more expensive compared to the Octopus implementation when the spectrum has to be obtained over a large energy range. At present, it is the most accurate method to calculate electronic spectra of large PAHs down to wavelengths of ca. 200 nm, i.e., for energies smaller than 6.2 eV (5 $\mu$m$^{-1}$). We focused on seven selected PAHs with compact structures, possibly representing the average PAH population of the ISM. As indicated by astronomical observations in the mid-IR, interstellar PAHs are rather compact than elongated \citep{leger89}. The B3LYP functional \citep{stephens94, becke93} in conjunction with the 6-31+G(d) basis set was used to optimize the ground state geometries and to calculate the vertical electronic transition energies afterward. Additional vibrational excitations cannot be treated in this approach. However, at higher energies, a substantial band broadening is expected, resulting in spectra without discernible vibrational pattern even for cold matrix-isolated or gas phase molecules (see, e.g., Section \ref{results}).

\begin{figure}[t]\begin{center}
\epsscale{1.15} \plotone{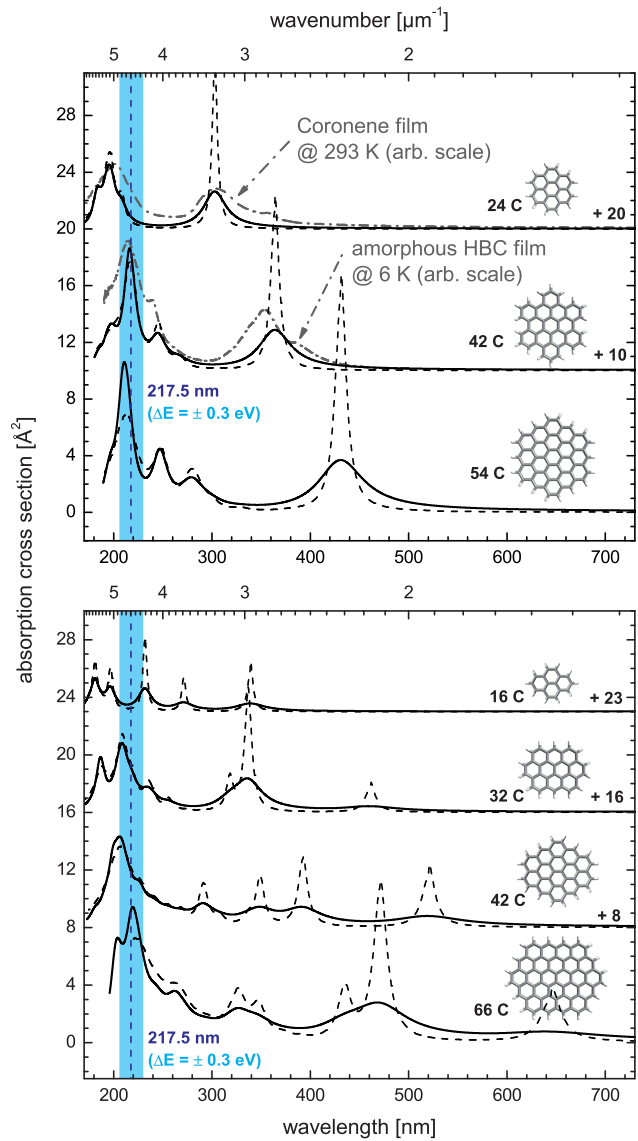} \caption{Theoretical electronic absorption spectra of different compact PAHs with D$_{\text{6h}}$ (top panel) and D$_{\text{2h}}$ (bottom panel) symmetry calculated at the B3LYP/6-31+G(d) level of theory. A convolution using Lorentzians with energy-dependent widths has been applied to visually enhance the weaker bands at longer wavelengths (dashed lines). For comparison, the absorption spectra of thin films of coronene and HBC are shown (dash-dotted lines). The position of the interstellar UV bump is marked at 217.5 nm, plus-minus a reasonable 0.3 eV uncertainty accounting for the inaccuracy of the TD-DFT calculation.} \label{fig1}
\end{center}\end{figure}

Figure \ref{fig1} displays the calculated electronic absorption spectra of compact PAHs belonging to the D$_{\text{6h}}$ (top panel) and D$_{\text{2h}}$ (bottom panel) point groups. It should be taken into account that the uncertainty for the band positions at this level of theory can typically reach up to 0.3 eV \citep[see, e.g.,][]{hirata99}. At the moment, we limit the discussion to neutral species, as the ions demand higher computational efforts due to their open-shell structure. For each series (D$_{\text{2h}}$ and D$_{\text{6h}}$), the same symmetry-related selection rules apply, allowing direct comparison among the differently sized species. The theoretical stick spectra (oscillator strength vs. energy, not shown) have been convoluted with Lorentzian functions with full width half maximum (FWHM) of 3000 cm$^{-1}$ to obtain the spectra shown by solid lines. Similar absorption spectra can be expected for weakly bonded PAH clusters where vibrational patterns are smoothed out due to the van der Waals interactions among the molecules. To illustrate this point, the absorption spectra of thin films of coronene and HBC are shown for comparison (dash-dotted lines). For the spectra represented by dashed lines, the FWHM for each transition was chosen to be proportional to the number of energetically lower-lying transitions while keeping the area of each band proportional to the corresponding oscillator strength. The purpose of this convolution is to reflect the increased lifetime broadening at higher energies and to visually enhance the weaker bands at longer wavelengths. This also resembles more closely the situation of matrix-isolated molecules as will be seen in Section \ref{results}. However, one should consider that the band intensity (integrated cross section) of real molecules in the visible is usually distributed over several vibrational bands reducing the peak intensities.

Looking at both PAH series, a strong UV peak around 217.5 nm for molecules with more than $\approx$ 32 C atoms (ovalene) is apparent which is in agreement with the results of \citet{malloci04} who showed that the cations and anions also display this UV peak (see also Fig. \ref{fig2}). From the computational point of view, the transitions of large PAHs that produce the UV bands around 217.5 nm are almost not affected when one electron is removed from the highest occupied orbital or added to the lowest unoccupied orbital. Differences in the UV spectra of different charge states should mainly concern the bandwidths. This problem cannot be assessed with theoretical calculations and will be addressed in Section \ref{results}. While the feature around 217.5 nm seems to be a common property among different sufficiently large PAHs, bands at longer wavelengths depend more strongly on the specific molecular structure. The averaging over several PAH molecules will produce a rather smooth extinction curve with a broad UV bump, possibly around 217.5 nm, as main feature. Even though the species we chose for the calculations are not large enough (due to computational limitations), a shift of their UV bump to even longer wavelengths, i.e., beyond 220 nm, can be expected for increasing molecular sizes. It is known that also amorphous graphitic material and nanoparticles, usually containing both aromatic and aliphatic components, can feature a similar UV bump in the wavelength range between 220 and 260 nm \citep{schnaiter96, schnaiter98, llamas-jansa07}. The varying band position might, at least partly, be related to the average size of the aromatic planes. However, the width of the bump in these nanoscopic and macroscopic materials is often quite broad due to several possible effects, such as particle agglomeration, van der Waals and covalent bonding between the aromatic planes, or just broad size distributions of the aromatic planes.

\subsection{FUV Rise}
In the wavelength range below 200 nm, the spectra of large PAHs share further characteristics. These characteristics can only be obtained by theory, as almost no experimental data are available. For the calculation of the FUV absorption bands of PAHs, the method described in the previous section is computationally too expensive due to the high density of states in this energy regime. The determination of electronic transitions should now be based on the TD-DFT implementation of the Octopus software package \citep{castro06,marques04}. Compared to the calculations by \citet{malloci04}, we chose slightly different parameters. The Octopus code relies on numerical meshes instead of basis sets to calculate the wave functions in real-space and real-time. Followed by an initial very short electromagnetic pulse, exciting all frequencies of the system, the time-dependent dipole moment is calculated from which the linear optical absorption spectrum, comprising all possible electronic bound-bound transitions, can be derived. The crucial parameters of the calculation are the functional (here B3LYP), the volume of the numerical box (sphere around each atom with radius 3 \AA), the grid spacing (0.25 \AA), the time integration length (20 $\hbar$/eV), and the time step (0.002 $\hbar$/eV).

Here, we only want to emphasize the most important and securest information that can be extracted from these calculations and that concerns the compatibility of the FUV absorption properties of large PAHs with the interstellar extinction. A more extensive discussion can be found in the publication by \citet{cecchi08}. In this paper, different FUV rises that are observed for the interstellar extinction in different lines of sight were interpreted to stem from different ionization states in the same PAH population. The calculated PAH spectra showed a tendency for a reduction of the absorption cross section in the gap between the highest-energy $\pi$ -- $\pi^*$ transitions and the $\sigma$ -- $\sigma^*$ FUV rise when going from anions over neutral molecules to cations. However, this effect seems to be more pronounced in smaller PAHs and the majority of the 40 different molecules in this model contained less than 32 C atoms.\footnote{For details, see the online database at http://astrochemistry.ca.astro.it/database/pahs.html.} Also, no physical explanation was given for this behavior. 

\begin{figure}[t]\begin{center}
\epsscale{1.2} \plotone{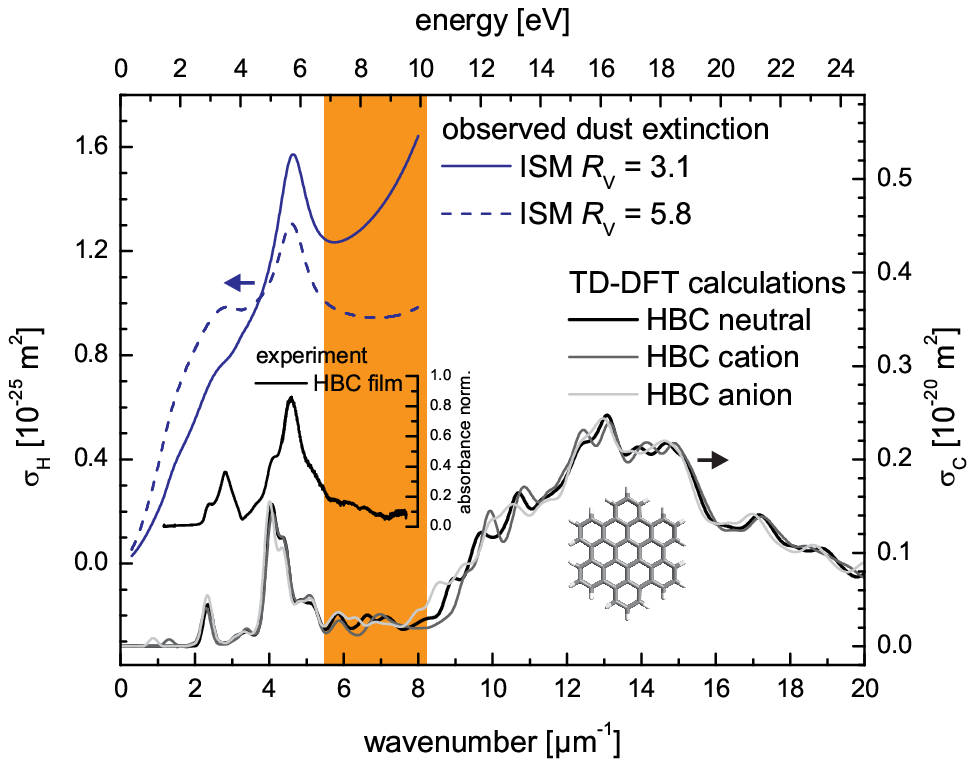} \caption{TD-DFT (B3LYP) spectra of neutral and ionized hexabenzocoronene (HBC) calculated with the Octopus code \citep{castro06,marques04} for the elucidation of the FUV properties of large PAHs. The shaded area roughly marks the absorption gap between the $\pi$ -- $\pi^*$ bands and the onset of the $\sigma$ -- $\sigma^*$ hump. Note that the energetic positions of the $\pi$ -- $\pi^*$ resonances below 6 $\mu$m$^{-1}$ are actually underestimated by this method which becomes apparent in comparison with the measured spectrum of a film-like deposit of HBC \citep{steglich10}. The mean interstellar extinction curves for two different reddening parameters $R_{\text{v}}$ (where $R_{\text{v}}=3.1$ is the interstellar average) are shown for comparison \citep{cardelli89}. The $y$-axes are labeled in units of cross section per H atom $\sigma_{\text{H}}$ or C atom $\sigma_{\text{C}}$, respectively.} \label{fig2}
\end{center}\end{figure}

Using the Octopus code, we calculated the spectra of neutral, cationic, and anionic HBC which are displayed in Figure \ref{fig2}. The spectra are very similar to the ones obtained by \citet{malloci04} who used the LDA (local-density approximation) functional. The widths of the absorption bands are purely artificial and depend solely on the integration length used in the calculation. The area of each band, however, is directly related to the oscillator strength of the corresponding transition(s). The shown spectra of HBC below 8 $\mu$m$^{-1}$ are dominated by $\pi$ -- $\pi^*$ transitions. The calculated energies of their band positions, however, are somewhat too low as becomes obvious upon comparison with the measured spectrum of a film-like HBC deposit. Also, the calculated strength of the transition at 2.3 $\mu$m$^{-1}$ is underestimated by a factor of two. Note that the broad bands of the HBC film will become much narrower if the molecules are isolated from each other in the gas phase or in a low-temperature matrix. By comparing with experimental gas phase spectra of anthracene, it was shown that the general trend of the broad $\sigma$ -- $\sigma^*$ hump above 8 $\mu$m$^{-1}$ as calculated with the Octopus code is in good agreement with measurements \citep{malloci04}. In the energy range above 2 $\mu$m$^{-1}$, the calculated spectra of differently charged HBC molecules are principally very similar to each other as this range is dominated by transitions where electrons are promoted from occupied to unoccupied molecular orbitals, i.e., electrons overleap the band gap (B- and C-type transitions). Further to the red ($<$ 2 $\mu$m$^{-1}$), additional transitions to the semi-occupied orbital of the cation (A-type transitions), as well as combined transitions to and from the semi-occupied orbital of the anion, give rise to additional bands which are absent for the neutral molecule.

For the interstellar extinction, no correlation could be observed between the bump at 217.5 nm and the non-linear FUV rise at shorter wavelengths \citep{cardelli89}. In order to ascribe the UV bump to a PAH feature, the FUV absorption properties of PAHs have to be compatible with the observed FUV rise of interstellar dust which is weakest for lines of sight with strong reddening (see Figure \ref{fig2}). As is apparent from the calculated HBC spectra, the onset of the $\sigma$ -- $\sigma^*$ hump is at rather high energies (8 $\mu$m$^{-1}$). The absorption gap between the strongest $\pi$ -- $\pi^*$ resonance and the $\sigma$ -- $\sigma^*$ absorption onset fits quite well to the observed FUV extinction properties of the interstellar dust. Furthermore, the theoretical spectra illustrate that the effect which was described by \citet{cecchi08}, i.e., the differently pronounced gap for different charge states, does not necessarily have to be true for larger species. Certainly, alternative scenarios that could explain the observed variations of the strength of the non-linear FUV rise, such as changes in the composition of other dust components, especially aliphatic components in carbonaceous grains, have to be explored. %In our opinion, the varying strong FUV rises of different astronomical sightlines are more likely predominantly caused by changes in the composition of other dust components, especially aliphatic components in carbonaceous grains, rather than by changes in the charge state of large interstellar PAHs.

\section{Laboratory-measured spectra of selected neutral and ionized PAHs} \label{results}

\subsection{Matrix Isolation Spectroscopy}
\begin{figure}[t]\begin{center}
\epsscale{1.15} \plotone{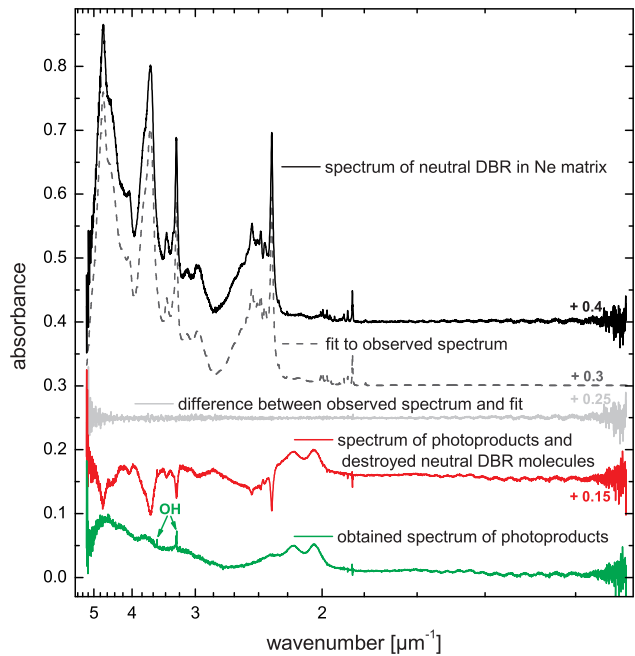} \caption{Procedure for obtaining the full-range spectrum of ionized PAH molecules, here DBR, in an inert gas matrix is illustrated (see the text for an explanation). The spectrum of neutral DBR shown at the top was baseline-corrected to account for light scattering from the matrix. The finally obtained spectrum of ionized DBR at the bottom displays an uncorrected interference pattern in the VIS. This has been removed in the spectrum shown in Fig. \ref{fig3}.} \label{fig3a}
\end{center}\end{figure}

This section addresses the question how photoionization influences the UV absorption bands of large PAHs, especially the bump around 217.5 nm. According to the calculations presented before and the theoretical studies by \citet{malloci04} the energies and strengths of the short-wavelength B- and C-type transitions of large neutral PAHs ($\lambda <$ 450 nm in the case of HBC) are almost unaffected upon removal or attachment of one electron. Potential differences, e.g., concerning the band shapes and positions, however, are only accessible via laboratory measurements. For this purpose, we used matrix isolation spectroscopy (MIS) to measure the wavelength-dependent photoabsorption between 190 and 850 nm of HBC and DBR molecules isolated in solid Ne at 7 K. The number of Ne atoms relative to the number of molecules to be investigated (isolation ratio) was on the order of $15000$ for DBR and $30000$ for HBC, respectively. After preparation of the matrix doped with the neutral PAH molecules, the sample was exposed to the FUV radiation (10.2 -- 11.8 eV) of a hydrogen-flow discharge lamp \citep{warneck62} to create ions from the neutral parent molecules. During the irradiation of typically 15 min, the FUV intensity on the sample surface was roughly $10^{15}$ -- $10^{16}$ photons m$^{-2}$ s$^{-1}$. The applied low FUV doses, as well as the inability of the molecules to diffuse through the matrix exclude the possibility that more complex chemical alterations than simple one-photon induced reactions (here only ionization) take place. The setup for MIS and FUV processing is described in more detail elsewhere \citep{steglich11}. Parts of the spectrum of ionized HBC in Ar matrix, which have been chosen for presentation because of a better signal-to-noise ratio, were recorded with a different setup \citep{bouwman09}. Due to the close proximity of the molecules in the matrix, recombination reactions between already ionized molecules and released electrons limit the maximum ion yield to values somewhere below 20\% after a certain FUV dose is reached. To disentangle the absorption bands of ionic and neutral species which, especially in the UV range, overlap each other, the following procedure was performed (see Fig. \ref{fig3a}). The transmission spectrum measured after irradiation was first baseline-corrected against the spectrum before irradiation, i.e., the spectrum of the neutral molecules. The spectrum thus derived in units of absorbance (red curve in Fig. \ref{fig3a}) displays positive features from species created during the irradiation (ions) and negative features from species that were destroyed (neutrals). Then, the contributions of the destroyed neutral molecules were simply subtracted by using an accurate fit obtained from the complete spectrum before irradiation. After subtraction, the spectrum contains only bands from species that were created during the irradiation. Actually, this procedure works quite well in the UV range despite the strong overlap of bands of the neutral and ionized molecules. For instance, even the sharp lines at 308 nm (3.25 $\mu$m$^{-1}$) and 283 nm (3.53 $\mu$m$^{-1}$) from the OH radical (green curve in Fig. \ref{fig3a}) which was created by photodissociation of trace amounts of water are properly reproduced. Problems can only occur in spectral ranges where the spectrum of the neutral precursor contains sharp bands that are represented by only one or two data points. For DBR, this is the case only at 540 nm (1.85 $\mu$m$^{-1}$).

The electrons released from the neutral parent molecules are usually trapped at defects and impurities in the matrix. Therefore, cations are generally created in excess compared to anions \citep[see, e.g.,][]{salama91,salama93}. The ratio between cation and anion number densities depends on several parameters, such as the purity and crystallinity of the matrix, the isolation ratio, or the electron affinities of the molecules and impurities. The possible contribution of anions to our spectra will be discussed in the following subsections.

\subsection{Dibenzorubicene (C$_{30}$H$_{14}$)}
\begin{figure}[t]\begin{center}
\epsscale{1.15} \plotone{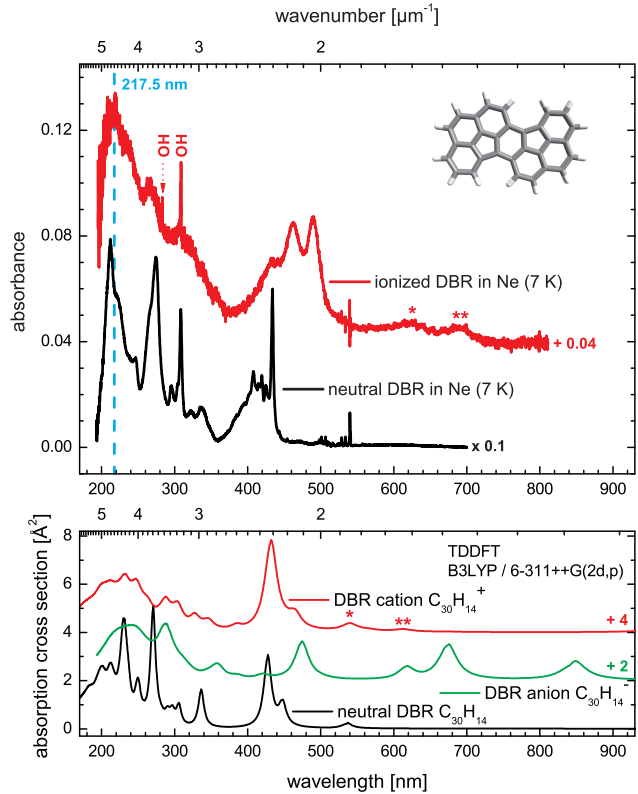} \caption{Top: measured spectra of matrix-isolated neutral and ionized dibenzorubicene (C$_{30}$H$_{14}$, DBR) at low temperature. Sharp features from the OH radical created by the dissociation of trace amounts of water are marked accordingly. The position of the interstellar UV bump is indicated by the vertical dashed line. Bottom: calculated spectra of DBR (B3LYP/6-311++G(2d,p)) in different charge states.} \label{fig3}
\end{center}\end{figure}

The measured spectra of neutral and ionized matrix-isolated DBR are displayed in the top panel of Figure \ref{fig3}. Detailed information on the spectroscopy, both in absorption and emission, of neutral DBR is given in another paper \citep{rouille10}. DBR can be regarded as a planar fragment of the fullerene C${_{70}}$ (with H-saturated dangling bonds), which was recently discovered in the planetary nebula Tc 1 \citep{cami10}. The neutral molecule has several strong absorption peaks in the UV, one specific at 212 nm in close proximity of the 217.5 nm interstellar bump. Therefore, it is an ideal test molecule to investigate how the high-energy $\pi$ -- $\pi^*$ transitions are affected by ionization.

The theoretical spectra of neutral, cationic, and anionic DBR are presented in the bottom panel of Figure \ref{fig3}. They were obtained in the same manner as the dashed spectra of Figure \ref{fig1}, i.e., using the TD-DFT approach of the Gaussian09 program, applying the 6-311++G(2d,p) basis set, and performing a convolution with Lorentzians, where the widths were proportional to the number of energetically lower-lying states. The optimized ground state geometry of DBR in all three charge states was found to be C$_{2\text{h}}$. The electronic ground state symmetries are $^1$A$_{\text{g}}$ (neutral), $^2$B$_{\text{g}}$ (cation), and $^2$A$_{\text{u}}$ (anion). For the neutral and the cation, electronic transitions from the ground state to excited states of A$_{\text{u}}$ and B$_{\text{u}}$ symmetry are dipole-allowed. For wavelengths shorter than 500 nm, the electronic excitation schemes of the various transitions are very similar for the neutral and the cation. Computational differences mainly arise from the broader cation bands which are caused by the convolution procedure and the fact that the cation has several low-energy transitions unlike the neutral molecule. Above 500 nm the theoretical spectrum of the DBR cation is dominated by weak transitions where electrons are promoted to the semi-occupied orbital, e.g., D$_5$(B$_{\text{u}}$)$\leftarrow$D$_0$(B$_{\text{g}}$) at 673 nm, D$_6$(B$_{\text{u}}$)$\leftarrow$D$_0$(B$_{\text{g}}$) at 613 nm, D$_8$(B$_{\text{u}}$)$\leftarrow$D$_0$(B$_{\text{g}}$) at 567 nm, and D$_{10}$(B$_{\text{u}}$)$\leftarrow$D$_0$(B$_{\text{g}}$) at 540 nm. Two further dipole-allowed (B$_{\text{u}}$) and two dipole-forbidden (A$_{\text{g}}$) transitions are predicted beyond 900 nm, outside of the accessible scanning range. The symmetry-related selection rules that apply to the anion are somewhat different. Due to its \textit{ungerade} ground state, dipole-allowed transitions are only possible to A$_{\text{g}}$ and B$_{\text{g}}$ states. Therefore, its calculated spectrum deviates more strongly from the spectra of the other two species. In view of the experimental spectrum of ionized DBR and the lack of strong absorption bands beyond 600 nm, we conclude that much less anions were formed than cations and we ascribe the measured spectrum solely to the DBR cation.

Except for a sometimes wrong ordering of states, it can be seen that the applied TD-DFT method works fairly well in the case of the neutral molecule. The S$_{1}$(B$_{\text{u}}$)$\leftarrow$S$_0$(A$_{\text{g}}$) transition, for instance, is predicted at 536 nm, while the origin band is actually measured at 540 nm. Two close-lying bands, S$_{2}$(B$_{\text{u}}$)$\leftarrow$S$_0$(A$_{\text{g}}$) and S$_{4}$(B$_{\text{u}}$)$\leftarrow$S$_0$(A$_{\text{g}}$), are calculated to be at 448 and 427 nm. The weaker one at 448 nm (S$_{2}$), with observed origin at 434 nm, is responsible for the vibrational pattern on top of the broad band (S$_{4}$) that can be seen in the experimental spectrum between 360 and 440 nm. These two bands have their counterparts with similar band positions and strengths in the cationic molecule, where they appear broader and slightly red-shifted in the measurement. To the red of these two bands, two weak and broad features (marked with $^{*}$ and $^{**}$) are observed around 620 and 690 nm which could correspond to the D$_{10}$$\leftarrow$D$_0$ and D$_{6}$$\leftarrow$D$_0$ transitions.

In the wavelength range between 340 and 200 nm, more than 100 single electronic transitions are predicted by TD-DFT for the cation. Roughly 45 of them have non-zero oscillator strengths and actually contribute to the observed broad $\pi$ -- $\pi^{*}$ structure peaking at ca. 215 nm. Except for minor peak shifts, this band system resembles a broadened version of the analogous $\pi$ -- $\pi^{*}$ structure of the neutral precursor which essentially can be explained by similar excitation schemes of electronic C-type transitions for both molecules. Because of a stronger broadening for all UV bands, the bump peaking at 215 nm appears even more pronounced for the cation compared to the neutral.

\subsection{Hexabenzocoronene (C$_{42}$H$_{18}$)} \label{hbc_sec}
\begin{figure}[t]\begin{center}
\epsscale{1.15} \plotone{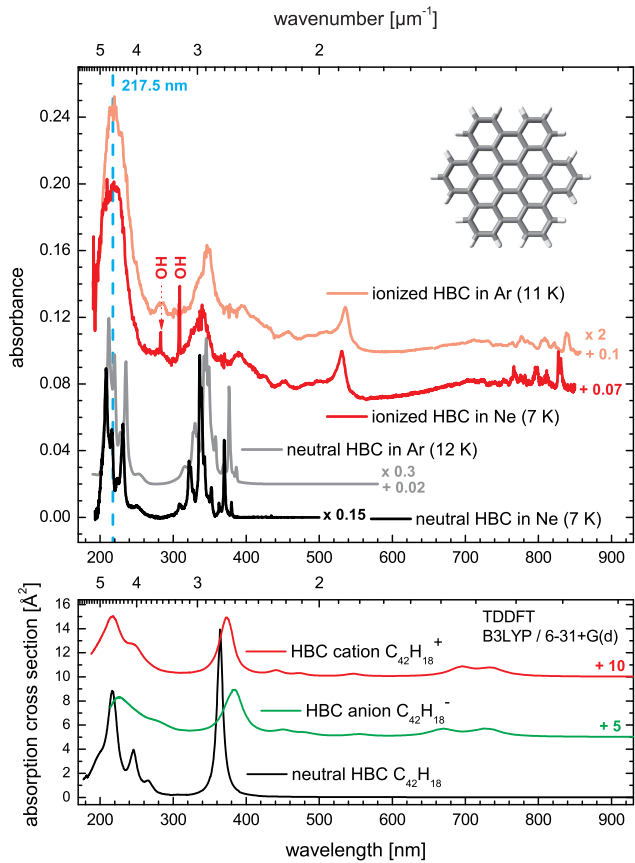} \caption{Top: measured spectra of matrix-isolated neutral and ionized hexabenzocoronene (C$_{42}$H$_{18}$, HBC) at low temperature. Sharp features from the OH radical created by the dissociation of trace amounts of water are marked accordingly. The position of the interstellar UV bump is indicated by the vertical dashed line. Bottom: calculated spectra of HBC (B3LYP/6-31+G(d)) in different charge states.} \label{fig4}
\end{center}\end{figure}

The spectroscopy of matrix-isolated neutral HBC is discussed in detail elsewhere \citep{rouille09}. Because of its all-benzenoid structure, it is believed that HBC is more abundant in space compared to other similarly-sized PAHs \citep{troy06}. The calculated spectrum of neutral HBC, obtained at the B3LYP / 6-31+G(d) level of theory, reveals a good match with the MIS measurements (see Figure \ref{fig4}), both in terms of relative band strengths and positions. Neutral HBC has a D$_{6\text{h}}$ ground state geometry. The intensity of its S$_{4}$(E$_{\text{1u}}$)$\leftarrow$S$_0$(A$_{\text{1g}}$) transition, calculated to be at 364 nm, is actually distributed over a complicated vibrational pattern that involves a$_{\text{1g}}$ modes of vibration and vibrational excitations belonging to an energetically lower B$_{\text{1u}}$ state, to which electronic transitions from the ground state are dipole-forbidden. Also the weak S$_{1}$(B$_{\text{2u}}$)$\leftarrow$S$_0$(A$_{\text{1g}}$) transition, for which a peak of an undefined vibrationally excited state has been measured at 434 nm, borrows some intensity from the S$_4$(E$_{\text{1u}}$) state. (Actually, the E$_{\text{1u}}$ state could also be S$_{3}$. For details, see \citet{rouille09}.) Quite well reproduced by the calculation, further strong bands are located around 217.5 nm.

The spectroscopic interpretation of the spectrum of ionized HBC in solid Ne is more complicated. The degenerate ground states of the HBC cation and anion cause Jahn-Teller interaction, which effectively leads to a reduction of the ground state geometry from D$_{6\text{h}}$ to D$_{2\text{h}}$. This also complicates the assignment of measured absorption bands on the basis of DFT calculations as these methods can already fail to correctly predict the distorted geometry of a degenerate ground state \citep[see, e.g.,][]{davidson83,russ04}. The comparison of the measured spectrum of ionized HBC with the TD-DFT calculated spectra  (bottom panel in Figure \ref{fig4}) does not permit to exclude the presence of anion bands, in contrast to the case for DBR. The band system with observed origin around 830 nm in the Ne matrix could have its theoretical counterpart in either cation or anion bands predicted around 740 nm. Using the real-time, real-space implementation of TD-DFT (see Figure \ref{fig2}), however, these anion bands are calculated to be much further to the red around 1100 nm (0.88 $\mu$m$^{-1}$), while the cation bands remain around 760 nm (1.31 $\mu$m$^{-1}$). As already stated, cation bands usually dominate the spectra of matrix-isolated photoionized PAHs. Therefore, we tentatively attribute all observed bands of the photoprocessed matrix (except the OH lines) to the HBC cation and discuss the results accordingly.

The electronic ground state of the HBC cation transforms according to the B$_{\text{1g}}$ irreducible representation of the D$_{\text{2h}}$ point group.\footnote{All given symmetries refer to the molecule being oriented in the $y$-$z$-plane ($x$=0). The $z$-axis traverses the C atoms and defines the direction in which the molecular structure is elongated.} Electronic transitions to $^2$A$_{\text{u}}$, $^2$B$_{\text{2u}}$, and $^2$B$_{\text{3u}}$ states are dipole-allowed. The theoretical spectrum contains several A-type transitions to the semi-occupied orbital at wavelengths longer than 500 nm. The lowest D$_{1}$(B$_{\text{2g}}$)$\leftarrow$D$_0$(B$_{\text{1g}}$) transition is actually calculated to be far beyond 2 $\mu$m. The observed band pattern between 750 and 835 nm may originate from the following calculated transitions: D$_{6}$(B$_{\text{3u}}$)$\leftarrow$D$_0$(B$_{\text{1g}}$) at 732 nm with oscillator strength $f=0.038$, D$_{7}$(A$_{\text{u}}$)$\leftarrow$D$_0$(B$_{\text{1g}}$) at 724 nm with $f=0.039$, and D$_{8}$(B$_{\text{3u}}$)$\leftarrow$D$_0$(B$_{\text{1g}}$) at 686 nm with $f=0.081$. Another transition D$_{9}$(A$_{\text{u}}$)$\leftarrow$D$_0$(B$_{\text{1g}}$) is predicted at 542 nm, close to the measured broad band with maximum at 531 nm. However, its strength ($f=0.025$), if it can be ascribed to the measured band, is considerably underestimated which, e.g., could be related to an underestimation of the Jahn-Teller distortion in the ground state by DFT.

At wavelengths shorter than 400 nm, the density of electronic transitions rises quickly. A detailed spectroscopic analysis is hardly practicable anymore. The comparison of ionized HBC in two different matrix environments (Ar and Ne) illustrates an important aspect concerning the absorption spectra of PAH cations in this wavelength range. As it was the case for DBR, the spectrum of ionized HBC contains broad bands which resemble broadened versions of analogue bands found in neutral HBC. Especially, the $\pi$ -- $\pi^{*}$ structure around 217.5 nm is even more apparent for the ionized molecule. If the strong band broadening could be ascribed to a matrix-induced effect, larger bandwidths for the Ar compared to the Ne environment would be expected, as interaction-related effects are usually more pronounced for atoms with higher polarizabilities. The shapes and widths of these features, however, do not seem to depend on the rare gas atoms surrounding the molecule (for $\lambda <$ 550 nm), indicating an intrinsic effect, e.g., very short lifetimes of the excited electronic states. On that supposition, low-temperature gas phase spectra of ionized medium-sized to large PAHs, such as DBR and HBC, at wavelengths shorter than approximately 400 nm would feature band shapes and widths similar to what has been measured for molecules isolated in the inert gas matrix. However, a small wavelength shift, usually to the red, has to be considered in that case. The matrix-induced redshift becomes obvious upon comparison of the Ne and Ar measurements. We will take a closer look at this in the following subsection.

\subsection{Implications for the Carriers of the Diffuse Interstellar Bands} \label{dib_sec}
A connection between large PAHs and the DIBs is often assumed. Nevertheless, DBR can be excluded as possible DIB carrier. While the case of the neutral molecule is discussed in another article \citep{rouille10}, cationic DBR does not exhibit sufficiently narrow bands - at least for wavelengths shorter than 900 nm. The first electronic transition of HBC in the gas phase causes some narrow and very weak bands between 410 and 435 nm (24390 -- 22990 cm$^{-1}$) not able to reproduce any DIB \citep[][]{kokkin08}. Stronger absorption bands appear further in the UV where no DIBs were observed. The HBC cation, on the other hand, displays a system of rather narrow bands between 750 and 850 nm. In Ne matrix, the strongest features were found at 830.8 nm (12037 cm$^{-1}$) and 827.3 nm (12088 cm$^{-1}$). The availability of the spectra of ionized HBC in matrices with different polarizabilities allows us to perform a linear extrapolation procedure to obtain the band positions in the gas phase. This procedure is described in another article \citep{gredel11}. The wavenumber position of a certain band in the gas phase $\tilde \nu_{gas}$ can be approximated via
\begin{equation}
\tilde \nu_{gas} = \tilde \nu_{Ne} + \frac{\tilde \nu_{Ne} - \tilde \nu_{Ar}}{\frac{\alpha_{Ar}}{\alpha_{Ne}} - 1} ,
\end{equation}
where $\tilde \nu_{Ne, Ar}$ are the respective band positions in the matrices. The polarizabilities amount to $\alpha_{Ar} = 1.64$ \AA$^3$ and $\alpha_{Ne} = 0.397$ \AA$^3$ \citep{radzig85}. The extrapolated gas phase positions and approximate oscillator strengths for bands of ionized HBC are summarized in Table \ref{table1}. For the bands between 750 and 850 nm (13400 -- 11800 cm$^{-1}$), the positions have been obtained by fitting the experimental spectra to Lorentzian functions. For the broad asymmetric band at 531 nm (18840 cm$^{-1}$), only the maximum position has been taken. We did not measure the absorption cross sections directly in the MIS experiments because the results are very often not accurate. Instead, the approximate oscillator strengths were determined by comparing the integrated band intensity with the integrated intensity of the strong band system between 200 and 260 nm. The accumulated oscillator strength of this band system was taken from the TD-DFT calculation ($f \approx 7$). The uncertainties of the determined $f$-values mainly reflect the expected inaccuracies of the calculated transition strengths. The inaccuracies were obtained for the neutral molecule upon comparison with experimental data of HBC in a solvent.

\begin{figure}[t]\begin{center}
\epsscale{1.15} \plotone{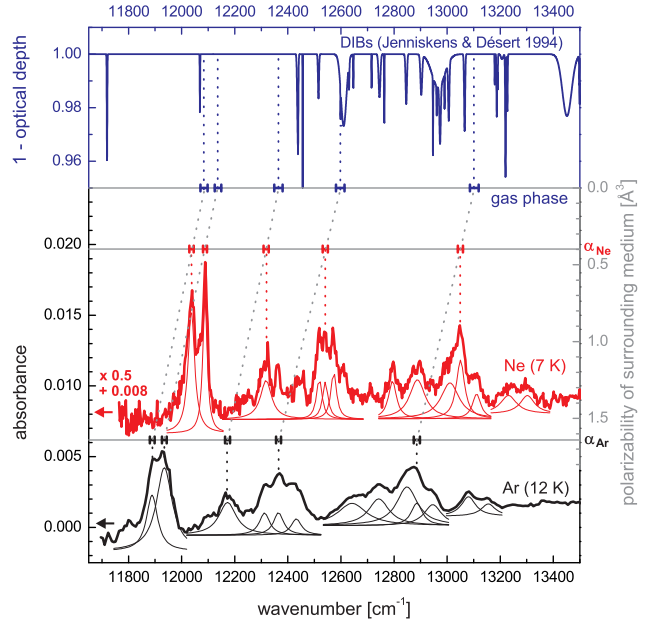} \caption{Comparison of absorption spectra of ionized HBC in Ar and Ne matrices with the spectrum of the diffuse interstellar bands \citep{jenniskens94}. Presumably, all measured bands are due to the HBC cation. The gas phase positions of the strongest HBC features which were obtained by a linear extrapolation procedure are indicated (see also Table \ref{table1}). The fitted Lorentzians were only used to determine the band positions.} \label{fig5}
\end{center}\end{figure}

\begin{figure}[t]\begin{center}
\epsscale{1.15} \plotone{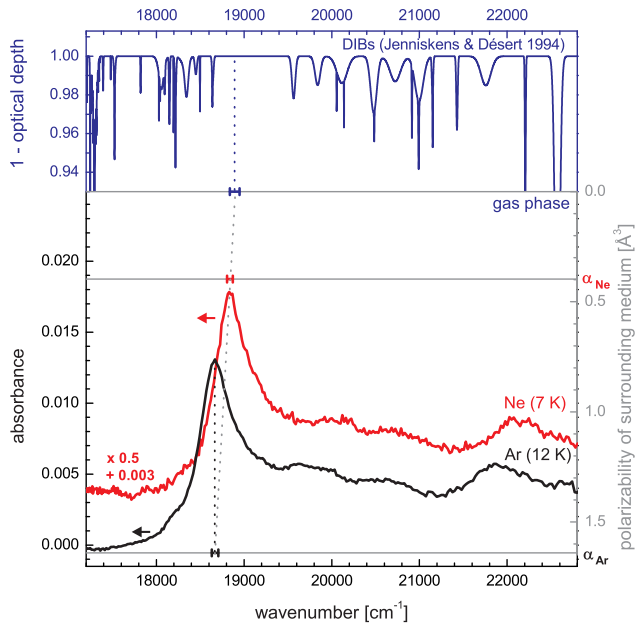} \caption{Comparison of absorption spectra of the HBC cation observed at 536 and 531 nm in Ar and Ne matrices with the spectrum of the diffuse interstellar bands \citep{jenniskens94}. The gas phase position (529 nm) was determined by a linear extrapolation.} \label{fig6}
\end{center}\end{figure}

The linear extrapolation procedure on the basis of the absorption spectra is furthermore illustrated in Figs. \ref{fig5} and \ref{fig6}. The bands that are displayed in Fig. \ref{fig5} may actually originate from more than one electronic transition which could also explain small differences in the matrix-induced wavelength shifts (see the third column in Table \ref{table1}). Moreover, these bands seem to be superimposed on a very broad band with a tail extending down to 600 nm (16670 cm$^{-1}$; see also Fig. \ref{fig4}). The spectra of Fig. \ref{fig6} probably display a vibronic progression with rather broad bands. The strongest feature is an asymmetric band peaking at 531 nm (18840 cm$^{-1}$) in Ne. Its band shape and bandwidth ($\approx$ 300 cm$^{-1}$) are identical in both matrix environments and are expected to be the same in the gas phase. A comparable band shape has been found at 462 nm in Ar matrix for the smaller coronene cation \citep{szczepanski93}, which has an electronic structure similar to the HBC cation.

The comparison of the band positions that were extrapolated for the gas phase with the synthetic DIB spectrum of \citet{jenniskens94}\footnote{The DIB spectrum has been calculated with the most recent data from the online catalog available at http://leonid.arc.nasa.gov/DIBcatalog.html.} demonstrates that ionized HBC can safely be excluded as a potential DIB carrier. Basically, the obtained gas phase positions and $f$-values could be used to search for interstellar features of HBC cations, but we are skeptical whether such an endeavor would be successful. A search for interstellar HBC in its neutral charge state was already unsuccessful \citep{gredel11}. We want to stress again that because of its size and superior stability (most stable PAH with 42 C atoms) one would expect to observe features of neutral or charged HBC on the interstellar extinction curve if the PAH hypothesis, i.e., the assignment of the DIBs to large PAHs, would be correct. However, in view of the expected chemical variety for PAHs of this size and the failure to detect HBC in interstellar spectra it seems unlikely that a few selected PAH molecules (out of several million possible structures) could be abundant enough to be responsible for one or more of the roughly 400 DIBs. Additionally, PAH spectra can be very complex with complicated vibrational and intermediate level structures which do not seem to be present in interstellar spectra. This is especially true for PAHs with molecular structures that are less symmetric than what is usually studied. Furthermore, electronic bands of cations appearing in the VIS can be very broad \citep[see, e.g.,][]{biennier03, sukhorukov04} making it impossible to connect them with the narrower DIBs. In summary, we would expect that the population of large interstellar PAHs produces a smooth extinction curve as it was already observed for mixtures of sufficiently large PAHs in inert gas matrices \citep{steglich10}.

\begin{table*}[tb]
\begin{center}\caption{Extrapolated Gas Phase Positions and Approximate Oscillator Strengths for Bands of Ionized HBC (see also Figs. \ref{fig5} \& \ref{fig6})} \label{table1}
\begin{tabularx}{0.7\textwidth}{ccccXc}
 \hline\hline
 \multicolumn{2}{c}{Band Position in Matrix [cm$^{-1}$]}& $\tilde \nu_{\text{Ne}} - \tilde \nu_{\text{Ar}}$ & Extrapolated Gas Phase && Approx. Osc.\\
 Ar (12 K) & Ne (7 K) & [cm$^{-1}$] & Position [cm$^{-1}$]   &&  Strength ($^{+20\%}_{-60\%}$) \\
 \hline
                          &                        &       &                          &&          \\
 11889 $\pm$ 9 & 12037 $\pm$ 8 & 148 & 12084 $\pm$ 14 && 0.0065 \\
                          &                        &       &                          &&          \\
 11935 $\pm$ 9 & 12088 $\pm$ 7 & 153 & 12137 $\pm$ 13 && 0.0032 \\
                          &                        &       &                          &&          \\
 12173 $\pm$ 9 & 12318 $\pm$ 10 & 145 & 12364 $\pm$ 16 && 0.0023 \\
                          &                        &       &                          &&          \\
 12312 $\pm$ 10 & 12518 $\pm$ 10 & 206 & 12584 $\pm$ 16 &\multirow{3}{*}{$\left.\begin{array}{c}\\\\\\\end{array}\right\}$}& \multirow{3}{*}{0.0115} \\
 12365 $\pm$ 10 & 12541 $\pm$ 10 & 176 & 12597 $\pm$ 16 &&  \\
 12432 $\pm$ 11 & 12573 $\pm$ 9 & 141 & 12618 $\pm$ 15 &&  \\
                          &                        &       &                          &&          \\
 12642 $\pm$ 11 & 12794 $\pm$ 9 & 152 & 12843 $\pm$ 16 && 0.0017 \\
                          &                        &       &                          &&          \\
 12743 $\pm$ 10 & 12888 $\pm$ 10 & 145 & 12934 $\pm$ 17 && 0.0034 \\
                          &                        &       &                          &&          \\
 12849 $\pm$ 15 & 13011 $\pm$ 15 & 162 & 13063 $\pm$ 25 &\multirow{2}{*}{$\left.\begin{array}{c}\\\\\end{array} \right\}$}& \multirow{2}{*}{0.0062} \\
 12886 $\pm$ 11 & 13050 $\pm$ 10 & 164 & 13102 $\pm$ 16 &&  \\
                          &                        &       &                          &&          \\
 12946 $\pm$ 11 & 13111 $\pm$ 11 & 165 & 13164 $\pm$ 18 && 0.0012 \\
                          &                        &       &                          &&          \\
 13080 $\pm$ 10 & 13231 $\pm$ 13 & 151 & 13279 $\pm$ 20 &\multirow{2}{*}{$\left.\begin{array}{c}\\\\\end{array} \right\}$}& \multirow{2}{*}{0.0028} \\
 13155 $\pm$ 10 & 13302 $\pm$ 13 & 147 & 13349 $\pm$ 20 &&  \\
                          &                        &       &                          &&          \\
                          &                        &       &                          &&          \\
 18670 $\pm$ 37 & 18840 $\pm$ 33 & 170 & 18894 $\pm$ 55 && 0.0589 \\
 \hline
\end{tabularx}
\end{center}
\end{table*}

\section{Conclusions} \label{conclusion}
We presented experimental and computational spectroscopic studies of electronic transitions in the UV-VIS of PAHs comprising up to 42 C atoms (66 C atoms in the calculations). Principally, these species share two common features in the UV. The first one is a strong $\pi$ -- $\pi^*$ resonance below 230 nm, which will be particularly pronounced for a mixture of (probably rather compact) PAHs and whose exact position will depend on the molecular mean size. It will probably peak around 217.5 nm if the size of the PAHs is such that they can be connected to the aromatic MIR emission bands. As was demonstrated for natural PAH mixtures \citep{ehrenfreund92, steglich10}, narrow bands below 400 nm appear on the collective $\pi$ -- $\pi^*$ slope of the absorption curve if sufficient amounts of small molecules ($\lesssim$ 22 C atoms) are present. However, sensitive searches failed to detect such sharp UV absorption features in the diffuse ISM \citep{clayton03, gredel11, salama11} indicating the absence or low abundance, respectively, of free-flying small PAHs which is in accordance with the size constraints of interstellar PAHs as inferred from the MIR observations. Stringent constraints on the abundance of specific PAHs have also been obtained by other studies \citep{kokkin08, rouille09, pilleri09}. At shorter wavelengths, the spectra are dominated by the broad $\sigma$ -- $\sigma^*$ hump whose properties are quite common, especially among larger molecules. Its absorption onset is around 125 nm (8 $\mu$m$^{-1}$) as was demonstrated for HBC. The gap between the highest-energy $\pi$ -- $\pi^*$ feature and the $\sigma$ -- $\sigma^*$ onset fits quite well to the observed interstellar extinction curve, even for lines of sight displaying a low FUV extinction.

Using HBC and DBR as examples, we investigated experimentally how photoionization influences the $\pi$ -- $\pi^*$ absorption bands for wavelengths shorter than 400 nm. It was found that the main effect is a broadening \footnote{The issue of the band broadening of the first electronic transitions and non-radiative relaxation rates in small neutral and ionized PAHs was recently summarized by \citet{pino11}.} in combination with slight wavelength shifts of the individual bands. The position of the complete $\pi$ -- $\pi^*$ structure around 217.5 nm remains rather unaffected which can be explained with similar excitation schemes of C-type transitions in this energy range for the neutrals and the cations. The spectra of ionized HBC in different inert gas matrices illustrate another important point. The widths and shapes of the broad bands at wavelengths shorter than 600 nm (in the case of HBC) do not depend on the specific matrix environment. These bands are much broader than what is expected from typical matrix-induced broadening effects indicating very short lifetimes of the excited states which are not entirely caused by the interaction with the matrix atoms. (Alternatively, other broadening mechanisms, like Franck-Condon vibronic broadening, may be considered \citep[see, e.g.,][]{sassara01}.) Therefore, absorption spectra of cold ionized HBC in the gas phase (below 600 nm) will most likely resemble the presented MIS spectra. However, a small shift of the band positions to shorter wavelengths can be expected.

If the interstellar UV bump at 217.5 nm is indeed a collective feature of the interstellar PAHs that are observed through their IR emission signatures, its observational constraints could be explained as follows. The varying width for different lines of sight might be an indication for different degrees of ionization. Alternatively, this could just hint at different size distributions of the molecules. (Also, both explanations could be possible.) Its seemingly fixed position can probably be explained by a more or less stable mean size of the PAHs in the mixture. Within the observational and computational limitations, it was already shown that the interstellar abundance of PAHs could suffice to explain the observed  bump strength \citep{cecchi08, steglich10}. As suggested by astronomical observations, the outflows of carbon-rich stars inject hydrogenated amorphous carbon (HAC) into the ISM \citep{mathis90}. Recent laboratory investigations of such materials indicate the formation of more aromatic compounds under long-term FUV light exposure and the appearance of a bump-like absorption feature around 217.5 nm \citep{gadallah11}. However, this feature is yet too broad in comparison with the interstellar UV bump, probably because the electronic $\pi$ resonances in the HACs are still affected by chemical and physical bonding between the aromatic subunits. Furthermore, these laboratory materials seem to need slightly more carbon than available for interstellar dust in order to reproduce the observed strength of the interstellar feature \citep{gadallah11} which may be rationalized by a high content of aliphatic structures. Further chemical alteration by, e.g., FUV irradiation or interstellar shocks in the ISM could lead to the destruction of such structures and the release of aromatic units, i.e., large PAHs, with sizes sufficient to explain the MIR emission as well as the UV bump and the smooth extinction curve below 400 nm. These large PAHs, either neutral or ionized, with variable hydrogen saturation, may then represent a more or less stable end product in the ISM.

Concerning the PAH hypothesis, we found no evidence that neutral and singly ionized cationic PAHs studied so far by us or other laboratory groups could be responsible for any of the DIBs. Especially, the case of HBC, which is rather large and by comparison very stable, is compelling. The apparent absence of electronic features of neutral and cationic HBC in the diffuse ISM suggests that, due to molecular diversity, the abundances of individual PAH species of this size are probably too low to allow detection in the UV-VIS spectral domain. Therefore, we conclude that the DIBs (at least the prominent ones) are most likely not signatures of the interstellar PAH family, if we consider only neutral and singly ionized cationic species with normal hydrogen saturation.\\
%As a consequence, the DIBs (at least the more intense ones) might not be signatures of the interstellar PAHs as structurally defined here\\
%Furthermore, this would imply that the DIBs (at least the more intense ones) cannot be caused by the interstellar PAH family.\\

This work was supported by the Deutsche Forschungsgemeinschaft (DFG, project no. Hu 474/21-2). The authors are grateful to Harald Mutschke and Kamel A. Khalil Gadallah for measuring the UV flux of the H$_2$ discharge lamp as well as to Hans-Joachim Kn\"{o}lker (TU Dresden) and Klaus M\"{u}llen (MPI for Polymer Research Mainz) for providing the DBR and HBC samples, respectively. M.S. thanks Ga\"{e}l Rouill\'e for his constant support through helpful discussions. We also thank the anonymous referee for suggestions that helped to improve the quality of the paper.\\

\end{document}